\documentclass[aps,prb,superscriptaddress,amsmath,amssymb,reprint]{revtex4-2}

\usepackage{graphicx}
\usepackage{dcolumn}
\usepackage{bm}
\usepackage[utf8]{inputenc}
\usepackage[T1]{fontenc}
\usepackage{mathptmx}
\usepackage[separate-uncertainty=true]{siunitx}
\usepackage{xcolor}

\usepackage{physics}
\providecommand{\mean}[1]{\expval{#1}}

\usepackage[pdfencoding=auto, psdextra]{hyperref}
\hypersetup{
    colorlinks,%
    citecolor=blue,%
    filecolor=blue,%
    linkcolor=blue,%
    urlcolor=blue
}

\begin{document}

\title[]{Electrical control of spin relaxation anisotropy during drift transport \\ in a two-dimensional electron gas}

\author{F. G. G. Hernandez}
    \email{felixggh@if.usp.br}
    \affiliation{Instituto de F\'{i}sica, Universidade de S\~{a}o Paulo, S\~{a}o Paulo, SP 05508-090, Brazil}

\author{G. J. Ferreira}
    \affiliation{Instituto de F\'{i}sica, Universidade Federal de Uberl\^{a}ndia, Uberl\^{a}ndia, MG 38400-902, Brazil}
    
\author{M. Luengo-Kovac}
    \affiliation{Department of Physics, University of Michigan, Ann Arbor, Michigan 48109, USA}
    
\author{V. Sih}
    \affiliation{Department of Physics, University of Michigan, Ann Arbor, Michigan 48109, USA}

\author{N. M. Kawahala}
    \affiliation{Instituto de F\'{i}sica, Universidade de S\~{a}o Paulo, S\~{a}o Paulo, SP 05508-090, Brazil}

\author{G. M. Gusev}
    \affiliation{Instituto de F\'{i}sica, Universidade de S\~{a}o Paulo, S\~{a}o Paulo, SP 05508-090, Brazil}
    
\author{A. K. Bakarov}
    \affiliation{Institute of Semiconductor Physics and Novosibirsk State University, Novosibirsk 630090, Russia}

\date{\today}

\begin{abstract}
Spin relaxation was studied in a two-dimensional electron gas confined in a wide GaAs quantum well. Recently, the control of the spin relaxation anisotropy by diffusive motion was first shown in D. Iizasa et al., arXiv:2006.08253 (2020). Here, we demonstrate electrical control by drift transport in a system with two-subbands occupied. The combined effect of in-plane and gate voltages was investigated using time-resolved Kerr rotation. The measured relaxation time present strong anisotropy with respect to the transport direction. For an in-plane accelerating electric field along $\left[110\right]$, the lifetime was strongly suppressed irrespective of the applied gate voltage. Remarkably, for transport along $\left[1\bar{1}0\right]$, the data shows spin lifetime that was gate-dependent and longer than in the $\left[110\right]$ direction regardless of the in-plane voltage. In agreement, independent results of anisotropic spin precession frequencies are also presented. Nevertheless, the long spin lifetime, strong anisotropy and drift response seen in the data are beyond the existing models for spin drift and diffusion. 
\end{abstract}

\maketitle

\section{Introduction}
A spin transistor is a device in which spin polarization can be transferred from source to drain while the spins orientation precesses around a momentum-dependent spin-orbit field \cite{Datta1990,Wolf2001,zutic2004}. In a device using a conductive channel with a two-dimensional electron gas (2DEG) hosted in a quantum well (QW), the application of a gate voltage allows the electrical control of the precession angle by tailoring the spin-orbit couplings (SOCs) \cite{Dettwiler2017}. Moreover, the magnitude and orientation of the spin-orbit field can be tuned by an in-plane electric field that adds a drift velocity to the 2DEG electrons \cite{kikkawa99,kato2004}.

It is essential for such processing devices to retain coherence during the transport of the information encoded in the spin polarization. Nevertheless, for nonballistic transistors \cite{Schliemann2003,egues2003,Ohno2008}, momentum scattering of electrons leads to precession around random spin-orbit fields, resulting in unwanted spin decoherence \cite{dyakonov}. One way to suppress this mechanism is to create an unidirectional effective field by the combination of the Dresselhaus and Rashba spin-orbit interactions (SOIs) with equal strength \cite{Bernevig2006, Schliemann2017}. In this case, a helical spin mode is formed with enhanced coherence time, termed persistent spin helix (PSH) \cite{Koralek2009,Walser2012}. Drift transport in those systems showed current-controlled temporal oscillations of the spin polarization due to the cubic SOI \cite{PRL2016}, which results in spin relaxation. Further increasing the drift velocities, the enhancement of the cubic Dresselhaus SOI was observed and related to a heated electron distribution \cite{PhysRevB.86.174301,Kunihashi2017,PhysRevB.99.125404,aipadvances2020}. Configurations for the PSH in spin Hall transistors have been also considered to overcome relaxation for drift-induced SOIs along the transport direction \cite{PhysRevB.101.155414}. 

Very recently, it was shown that the spin relaxation anisotropy can be controlled by diffusive motion \cite{arXiv200608253}, in complement to the extensive literature for stationary spins in GaAs systems \cite{Aniso1,Aniso2,Aniso3,Aniso4} including high mobility samples \cite{Aniso5,Aniso6}. Here, we demonstrate electrical control during drift transport in a 2DEG confined in a wide QW occuping two-subbands. For systems with more than one subband occupied \cite{Bernardes,Calsaverini,fu2015}, the additional degree of freedom introduces new characteristics to the PSH dynamics influenced by the intersubband scattering rates \cite{PhysRevB.95.125119}.  Also, a persistent skyrmion lattice was proposed in two-subband QWs employing orthogonal PSH layers \cite{fu2015,FuPRL}. Using optical techniques, we measured the spin relaxation time anisotropy under the combined effect of in-plane electric fields and gate voltages. For transport along $\left[110\right]$ (y), the relaxation time strongly decreases with the drift velocity but remains unaffected by the gate voltage. On the contrary, for drift along $\left[1\bar{1}0\right]$ (x), the relaxation time is less modified by the velocity but strongly depends on the gate voltage. This result, together with the independent determination of anisotropic spin precession frequencies, is presented considering the role of the spin-orbit interaction and possible mechanisms are discussed including intersubband effects. 

\section{Materials and Experiment}

\begin{figure}[ht]
    \centering
    \includegraphics[width=1\columnwidth,keepaspectratio]{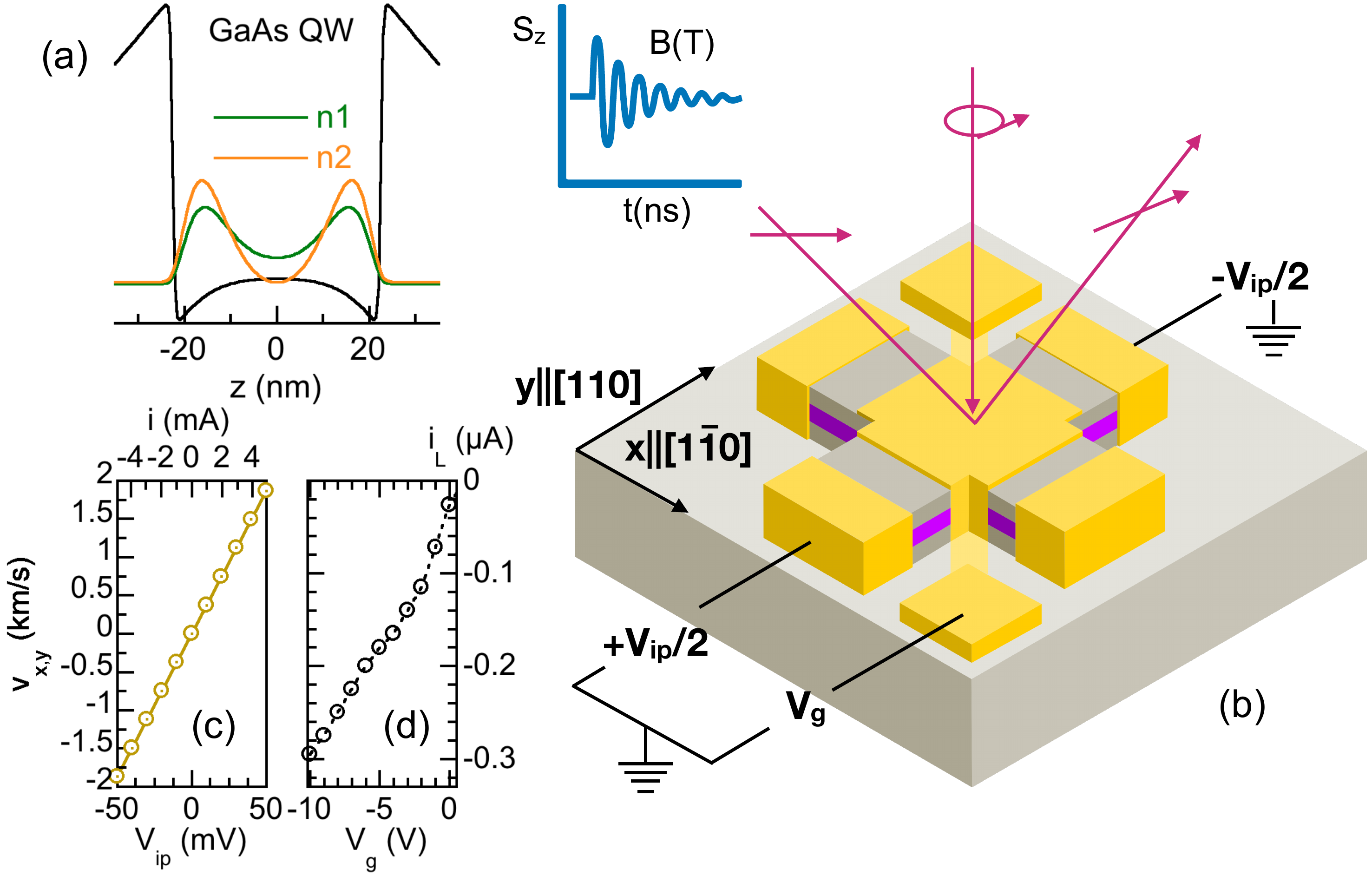}
    \caption{(a) Potential profile and the subbands charge density. (b) Device showing the connection scheme of $V_{\text{ip}}$ and V$_{\text{g}}$, and the geometry of the pump-probe experiment. (c) IV curves for the in-plane contacts and $\text{v}_\text{x,y}$. (d) Leakage current as function of V$_{\text{g}}$. 
   }
    \label{fig1}
\end{figure}

The studied sample is a \SI{45}{\nano\meter} wide GaAs QW, symmetrically doped with Si, and grown in the $\left[001\right]$ direction. The barriers were made of short-period AlAs/GaAs superlattices in order to enhance the mobility by shielding the doping ionized impurities. Similar multilayer systems have been extensively studied for the generation of current-induced spin polarization \cite{PRB2013,PRB2014,PRB2016,EPL2017}. Figure \ref{fig1}(a) shows the self-consistent solution of Schrödinger and Poisson equations for the QW band profile. Coulomb repulsion creates a soft barrier inside the well and configures the electronic system into symmetric and antisymmetric wave functions for the two lowest subbands.  The subband densities were found to be $3.7$ and $\SI{3.3e11}{\centi\meter^{-2}}$ from the Shubnikov-de Hass oscillations and the energy level separation is $\SI{2}{\milli\electronvolt}$.

A cross-shaped device (Figure \ref{fig1}(b)) was fabricated with channels along the $x$ and $y$ directions with a width of $\SI{270}{\micro\meter}$, where lateral contacts were deposited $l=\SI{500}{\micro\meter}$ apart to apply the in-plane voltages ($V_{\text{ip}}$). The drift velocity ($\text{v}_\text{x,y}$), shown in Figure \ref{fig1}(c), can be determined by tracking the spin packet displacement during a given time interval for a fixed in-plane electric field $E_{\text{ip}} = V_{\text{ip}}/l$ (see reference [\onlinecite{Luengo-Kovac2017}] for further details). The gate voltage (V$_{\text{g}}$) does not modify $\text{v}_\text{x,y}$ since the leakage current in Figure \ref{fig1}(d) is negligible in comparison with the drift current in (c). The out-of-plane electric field, proportional to the gate voltage, mainly modifies the Rashba SOC ($\alpha$) and only weakly changes the Dresselhaus cubic ($\beta_3$) and effective ($\beta^*=\beta_1-2\beta_3$) terms.

Optical pulses were used to inject and to detect the spin polarization dynamics. A simple scheme is shown in Figure \ref{fig1}(b). To perform time-resolved Kerr rotation (TRKR), a mode-locked Ti:sapphire laser with a repetition rate of $\SI{76}{\mega\hertz}$ and tuned to $\SI{816.73}{\nano\meter}$ was split into pump and probe beams. The pump beam was circularly polarized by means of a photoelastic modulator and the probe beam was linearly polarized. The polarization rotation of the reflected probe was recorded as function of the relative time delay ($t$) for a fixed external magnetic field ($B_\text{{ext}}$) of $\SI{0.2}{\tesla}$. The sample was rotated such that each channel was oriented parallel to $B_\text{{ext}}$ during measurement resulting in a drift-induced $B_\text{{SO}}$ always perpendicular to the external one. All measurements were performed at $\SI{10}{\kelvin}$.

\section{Results}

\begin{figure}[ht]
    \centering
    \includegraphics[width=1\columnwidth,keepaspectratio]{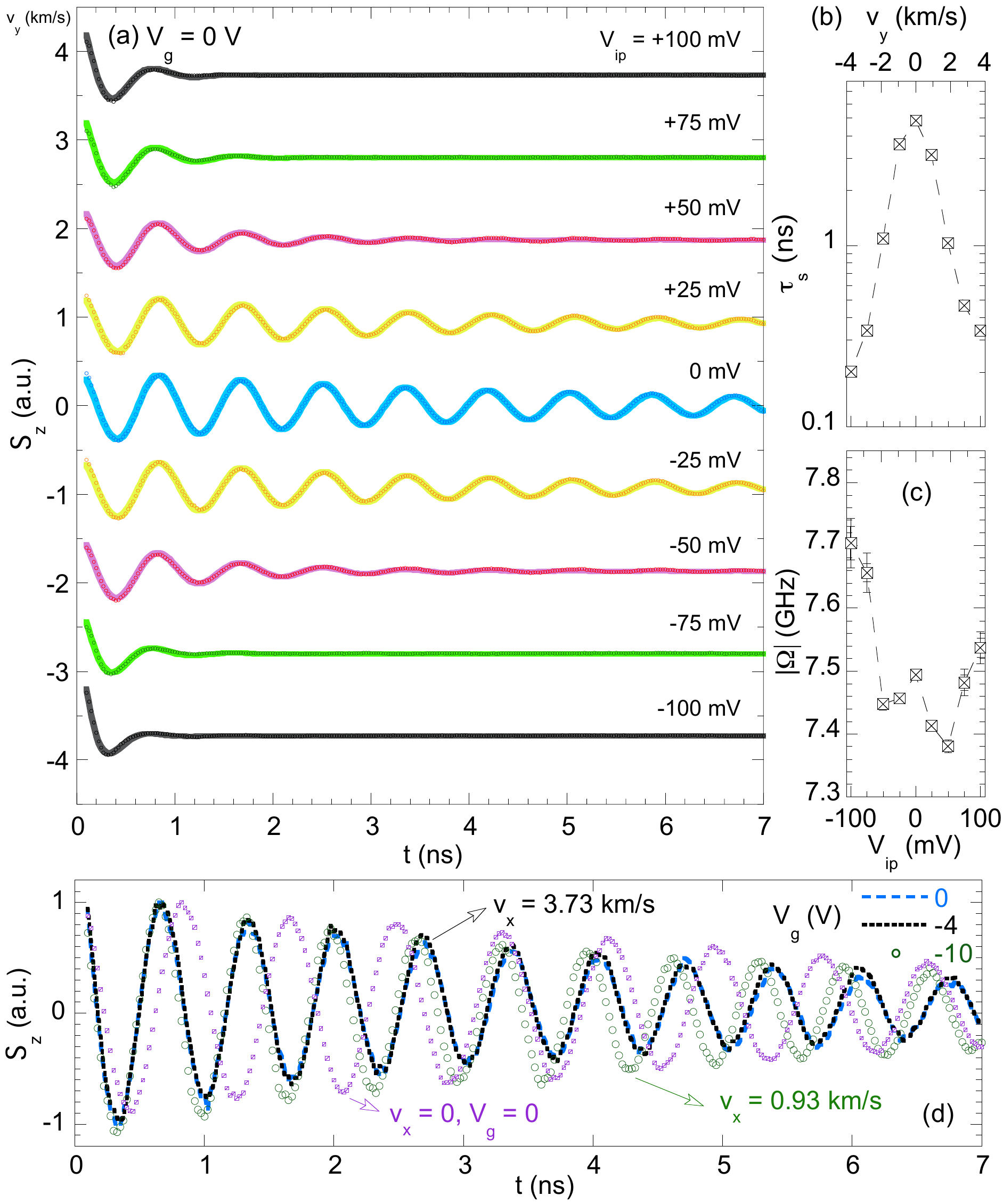}
    \caption{(a) TRKR scans for different velocities in the $y$ channel with zero gate voltage. The solid lines are data fittings from which spin relaxation times and precession frequencies were extracted and plotted in (b) and (c). (d) Scans with $\text{v}_\text{x}$ for several gate values.}
    \label{fig2}
\end{figure}

As commented above, the application of $E_{\text{ip}}$, in one of the device channels, adds a drift velocity to the 2DEG electrons. Thus, the measurement of the spin polarization at a fixed distance between pump and probe must consider the amplitude reduction due to the movement of the spin packet. We model the dependence of the Kerr rotation angle as follows: $S{z}(t) = A\cos(2\pi|\Omega_{x,y}| t)\times \exp(-t/\tau_s)\times \exp[-(\text{v}_{\text{x,y}}t)^2/4w^2]$, where $\Omega$ is the precession frequency, $\tau_s$ is the spin lifetime, and $w=\SI{23.5}{\micro\meter}$ is the width of the laser spot.
On this expression for $S_z(t)$, the drift correction arises from the motion of the gaussian packet (e.g., $\exp[-(x-\text{v}_xt)^2/2w^2]$) measured at $(x,y)=0$, and the factor 4, instead of 2, accounts for the addition of the probe beam with similar broadening $w$, which is assumed to be large, i.e. $w^2 \gg 2D_st$ within the time range of the measurements.
Note that the $\Omega$ and $\tau_s$ can be independently determined from the data fitting.

Figure \ref{fig2}(a) shows scans of TRKR for drift along $y$ with $V_{\text{g}}=0$. The spin oscillation is clearly damped when raising $V_{\text{ip}}$ from 0 up to $\SI{100}{\milli\volt}$. Fittings of $S_\text{z}$ are displayed as solid lines and the obtained parameters were plotted in Figure \ref{fig2}(b) and (c). As function of the drift velocity, the relaxation time presented a sharp decay by one order of magnitude and is accompanied by a precession frequency modification which are symmetrical for both polarities. Modifications of the electron g-factor by an in-plane electrical field have been also reported for (110)-oriented QWs \cite{chen} and in bulk epilayers \cite{marta}. On the contrary, for drift along $x$ and also in the absence of gate voltage (purple points and blue dashed curve in Figure \ref{fig2}(d)), the relaxation time is less affected by $V_{\text{ip}}$ with long-lived oscillations still found at $\SI{100}{\milli\volt}$ ($\SI{3.73}{\kilo\meter/\second}$). A drastic change on the precession frequency is observed for a nonzero velocity. These measurements for $V_{\text{g}}=0$ in Figure \ref{fig2}(a) and (d) clearly demonstrates the anisotropic action of a drift velocity on the spin relaxation. 

\begin{figure*}[ht]
    \centering
    \includegraphics[width=2\columnwidth,keepaspectratio]{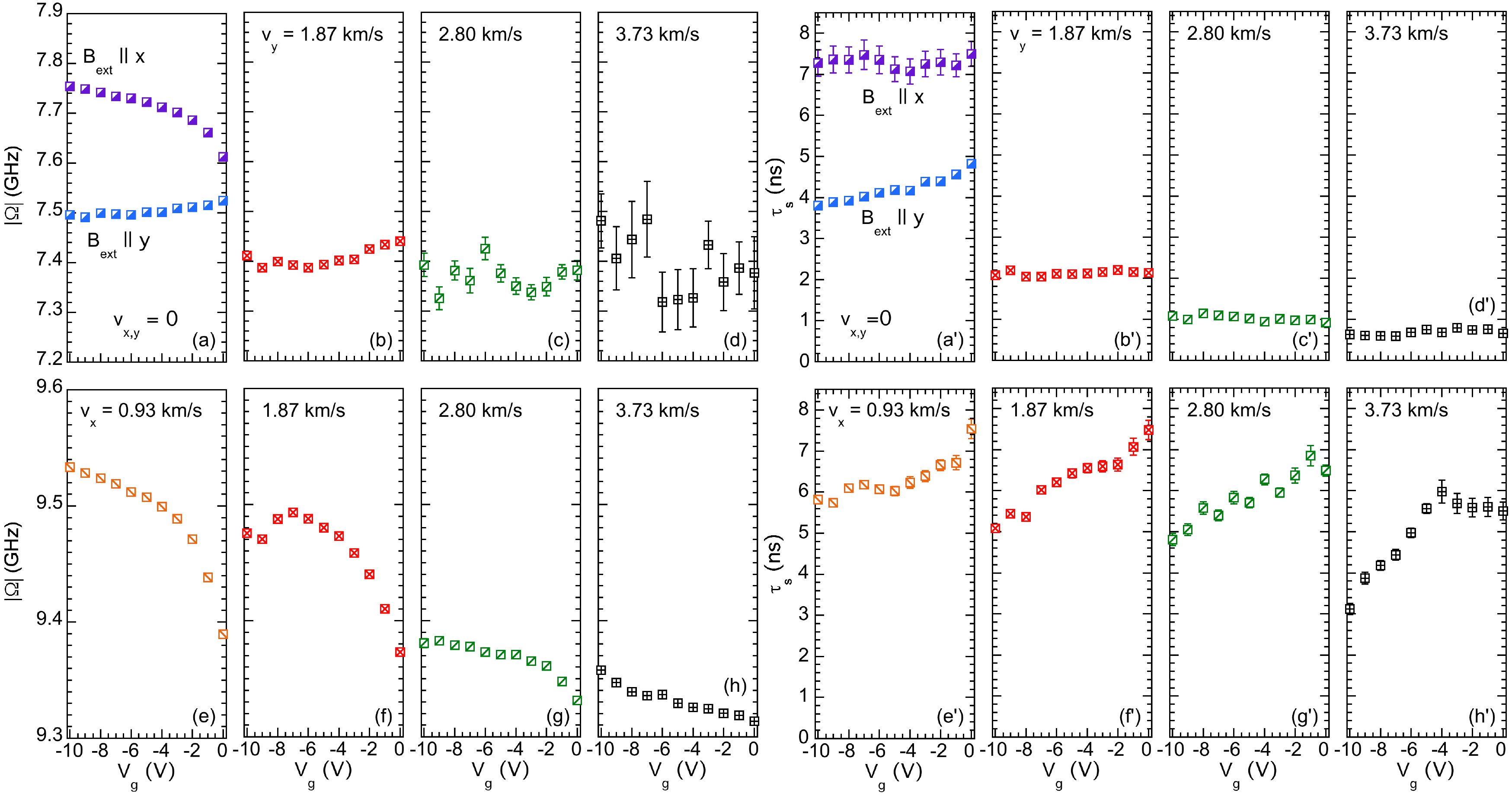}
    \caption{Gate dependence of the absolute value of the precession frequency (left) and spin relaxation time (right) for several drift velocities when the in-plane voltage is applied along $y$ in (a) to (d) or $x$ in (e) to (h) panels.}
    \label{fig3}
\end{figure*}

Furthermore, the effect of the gate voltage appears to be more complex in Figure \ref{fig2}(d). Keeping a fixed drift velocity, the signal amplitude increases by a moderate gate voltage at longer delay times (dashed black lines) that implies in an enhanced relaxation time. Also, large $V_{\text{g}}$ allows tuning the precession frequency. 

Repeating TRKR scans for several values of $V_{\text{ip}}$ and $V_{\text{g}}$, the combined effect of drift transport and gate control on the precession frequency (left) relaxation time (right) is plotted in Figure \ref{fig3}. Let's analyze first the data for zero drift velocity presented in Figures \ref{fig3}(a) and (a'). It is expected that $\Omega$ should be independent of $V_{\text{g}}$ for stationary spins and dominated by the external magnetic field. Nevertheless, Figure \ref{fig3}(a) displays an evident variation that is reduced when $V_{\text{g}}$ approaches zero. In a direct extrapolation, the curves cross at about +$\SI{1}{V}$ when $\alpha$ should be zero for the symmetric QW condition. The observed trend may indicate the role of a diffusive velocity that contributes to a finite spin-orbit field in the few mT range when $V_{\text{ip}}=0$. Such diffusive velocity must be considerably low due to the broad laser spot. For the relaxation time without drift velocity, the crossing showing such anisotropy removal is not as clear in Figure \ref{fig3}(a') and could indicate that the same anisotropy is more strongly displayed in the relaxation time than in the frequency. 

The precession frequency was also affected by the orientation and magnitude of the electrons motion as displayed in the left panel of Figure \ref{fig3}. For low velocities, the absolute value of $\Omega$ jumps to about $\SI{9.5}{\giga\hertz}$ for $\text{v}_x$ while remains in the range of $\SI{7.5}{\giga\hertz}$ for $\text{v}_y$ corresponding to effective g-factors of 0.54 and 0.42 in a field of $\SI{0.2}{\tesla}$, respectively. The change of $\Omega$ is surprising for drift along the external field direction as no contribution was expected when $B_\text{{SO}}$ points perpendicular to a much larger $B_\text{{ext}}$ \cite{henn}. A nonlinear modification of $\Omega$ was measured consisting of a sudden change at lower $V_{\text{g}}$. For larger $\text{v}_y$, strongly damped oscillations caused the uncertain determination of $\Omega$. However, for intermediate velocities as in Figure \ref{fig3}(b) and (f), we can clearly note an opposite trend of $\Omega$ with V$_{\text{g}}$ depending on the orientation of the drift transport and that leads to the anisotropy reduction for low gate voltage.

In the right panel, Figures \ref{fig3}(a')-(d') show the gate-control variation of the relaxation time for $\text{v}_y$. We observed a consistent reduction of the spin relaxation time with increasing drift velocities, as previously indicated in Figure \ref{fig2}(a) and (b), but there is no measurable dependence with the gate application. 

Moreover, for $\text{v}_x$, Figures \ref{fig3}(e')-(h') display a modification of the relaxation time on the order of $\SI{2}{\nano\second}$ in the studied V$_{\text{g}}$ range regardless of $\text{v}_\text{x}$. Surprisingly, a peak is formed for the largest velocity at intermediate gate voltages ($-4$ to $\SI{-6}{\volt}$) in Figure \ref{fig3}(h'). There could be two possibilities whether the spin relaxation mechanism depends on V$_{\text{g}}$ and $\text{v}_\text{x}$ such that the largest $\tau_s$ results for positive V$_{\text{g}}$ when $\text{v}_\text{x}$ is small and for V$_{\text{g}}\sim\SI{-4}{\volt}$ when $\text{v}_\text{x}=$ $\SI{3.73}{\kilo\meter/\second}$ or $\tau_s$ does not change with the velocity at this gate voltage. At the moment, it is not clear if this lifetime peak is moving from positive gate voltage towards this position as function of $\text{v}_\text{x}$ or if it appears due to the absence of relaxation for those V$_{\text{g}}$ values.

\begin{figure}[ht]
    \centering
    \includegraphics[width=1\columnwidth,keepaspectratio]{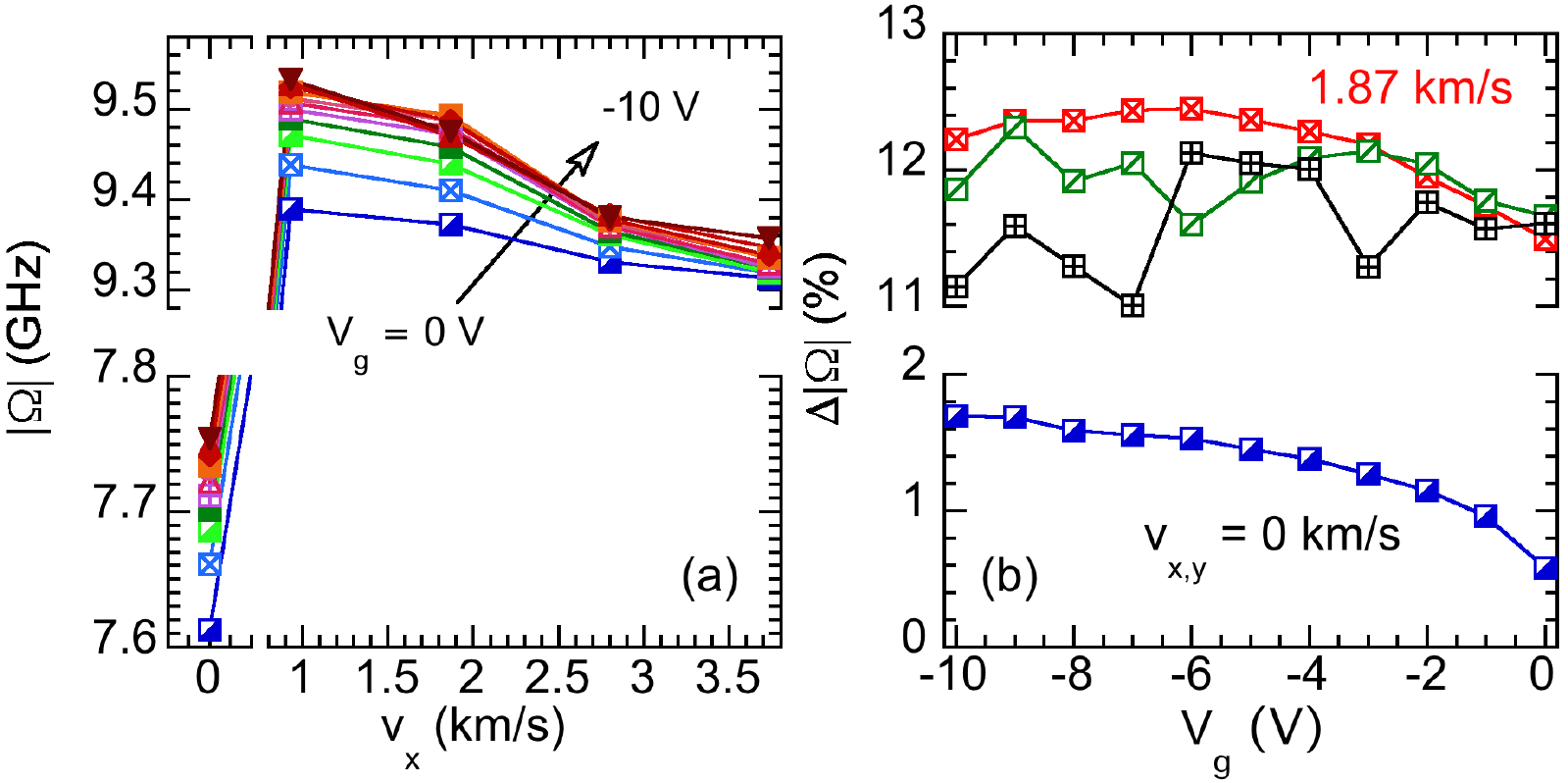}
    \caption{(a) Precession frequency for electrons moving with drift velocities along $x$ for several V$_{\text{g}}$ (see full color key in Figure \ref{fig5}). (b) Precession anisotropy as function of V$_{\text{g}}$ for different velocities.}
    \label{fig4}
\end{figure} 

Now, lets focus on a more detailed analysis of $\Omega$ and $\tau_s$ dependence with $\text{v}_\text{x}$ and in the characterization of the degree of electrical control for the anisotropy. Figure \ref{fig4}(a) shows the dependence of the precession frequency on $\text{v}_x$ for increasing V$_{\text{g}}$. All curves displayed an abrupt change for finite velocity, as discussed above, followed by a decreasing magnitude with increasing velocity. The anisotropy can be estimated by $\Delta|\Omega|=(|\Omega|_x-|\Omega|_y)/(|\Omega|_x+|\Omega|_y)$, where $|\Omega|_{x,y}$ is the magnitude of the precession frequency when the drift velocity is aligned with $x$ or $y$. 

It was found that $\Delta|\Omega|$ is in the order of 1\% for the probably low diffusive velocity and 12\% for several drift velocities as plotted in Figure \ref{fig4}(b). The variation with V$_{\text{g}}$ was limited to 1\% in all cases and the change was larger for low V$_{\text{g}}$, as already seen on the left panel of Figure \ref{fig3}.

\begin{figure}[h]
    \centering
    \includegraphics[width=1\columnwidth,keepaspectratio]{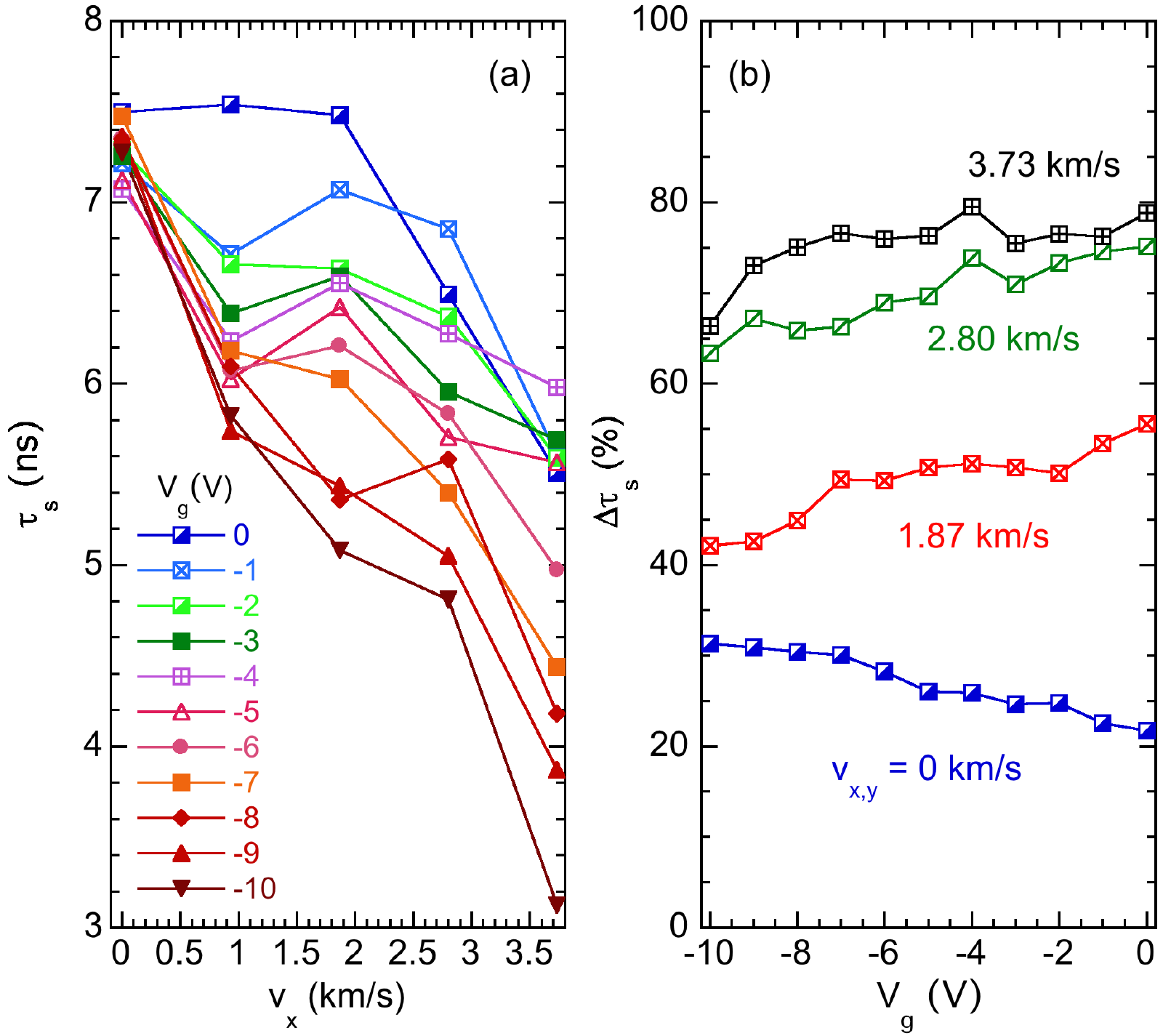}
    \caption{(a) Spin relaxation time for electrons moving with drift velocities along $x$ for several V$_{\text{g}}$. (b) Relaxation time anisotropy as function of V$_{\text{g}}$ for increasing $\text{v}_\text{x,y}$}
    \label{fig5}
\end{figure}

For the relaxation time, the anisotropy can be estimated in an analog way. As observed in Figure \ref{fig3}, $\tau_s$ for drift velocities along $y$ does not present a noticeable gate dependence. Therefore, the decay measured in Figure \ref{fig2}(b) is repeated for all V$_{\text{g}}$. Nevertheless, for drift velocities along $x$, $\tau_s$ displays a peculiar decay depending on V$_{\text{g}}$. The lower the magnitude of V$_{\text{g}}$, the greater is the range of drift velocities where we can found long-lived oscillations in Figure \ref{fig5}(a). 

Figure \ref{fig5}(b) shows that the effect of the electrons drift velocity on the relaxation anisotropy $[\Delta\tau_s=(\tau_s^x-\tau_s^y)/(\tau_s^x+\tau_s^y)$] is much stronger than the gate voltage action in the studied range and that gives a modification of about 10\%. The acceleration from zero to $\SI{3.73}{\kilo\meter/\second}$ increases $\Delta\tau_s$ from 20 to 80\% at V$_{\text{g}}=0$. Saturation was achieved around $\SI{2.8}{\kilo\meter/\second}$ approximately independent of V$_{\text{g}}$. 

\section{Discussion}

The experimental data presented above do not match the existing models for spin drift and diffusion in two-subband QWs \cite{FuPRL, PhysRevB.95.125119}. Indeed these references deal only with particular cases of a crossed PSH \cite{FuPRL}, and uncorrelated random walks between the subbands \cite{PhysRevB.95.125119}. In contrast, here the subband energy splitting is $\sim 2$~meV, while the spectral broadening due to scatterings is expected to be $0.6 < \hbar/\tau_0 < 6$~meV, for the momentum relaxation time $1 > \tau_0 > 0.1$~ps, which suggests a nearly degenerate regime where intersubband correlations might be relevant.

To illustrate how the current models fail to capture the results presented here, let us consider a generic spin relaxation model given by
\begin{align}
    \dfrac{\partial \bm{S}(t)}{\partial t}
    \approx 
    \begin{pmatrix}
        -\tau_x^{-1} & 0 & \mean{\Omega_y}
        \\
        0 & -\tau_y^{-1} & -\mean{\Omega_x}
        \\
        -\mean{\Omega_y} & \mean{\Omega_x} & -\tau_z^{-1}
    \end{pmatrix}\bm{S}(t),
\end{align}
where 
$\tau_x^{-1} = \tau_0\mean{\Omega_y^2}$,
$\tau_y^{-1} = \tau_0\mean{\Omega_x^2}$,
and $\tau_z^{-1} = \tau_x^{-1} + \tau_y^{-1}$ are the Dyakonov-Perel (DP) relaxation times, and $\mean{\bm{\Omega}}$ is average precession vector given by SOC and external $\bm{B}_{\rm ext}$ field. Here we are considering wide packets for simplicity ($w = 23$~$\mu$m), such that the spatial correlations vanish and the dynamics is simply that of a drifting packet with constant broadening $w \gg \sqrt{2D_st}$, i.e. $S_z \propto \exp[-(\bm{r}-{\rm \textbf{v}} t)^2/2w^2]$, with $\bm{r}=(x,y)$. Hereafter we neglect the spatial dynamics and focus on the spin precession and relaxation.

For $\bm{B}_{\rm ext} \parallel {\rm v} \parallel \hat{x}$ (and similarly for the $y$ direction), $\mean{\Omega_x} = g\mu_B B_{\rm ext}/\hbar \sim 7$~GHz as seen above for $B_{\rm ext}=200$~mT, while for the SOC obtained here, $\mean{\Omega_y} = g\mu_B B_{SO}/\hbar \sim 0.2$~GHz for $B_{SO} \sim 5$~mT at ${\rm v} \sim 3$~km/s \cite{FNBso}. Therefore, the effective field is dominated by the external $B_{\rm ext} = 200$~mT. On the other hand, the DP relaxation rate $\tau_s^{-1} \approx \tau_z^{-1}$ is dominated by the SOC, and to reach $\tau_s \approx 7$~ns we need $\tau_0 \sim 0.2$~ps. However, from the mobility we estimate $\tau_0 \rightarrow \tau_{\mu_e} = 76$~ps, and the finite temperature electron-electron scattering time $\tau_0 \rightarrow \tau_{ee}$ gives \cite{PhysRevB.86.174301} $10 > \tau_{ee} > 1$~ps for temperatures $10 < T_e < 60$~K. From Matthiessen's rule, $\tau_0^{-1} = \tau_{\mu_e}^{-1} + \tau_{ee}^{-1}$ we would get $1 < \tau_0 < 9$ ps, which is far from our $\tau_0 \sim 0.2$ ps estimate.

These numbers above show that $|\mean{\bm{\Omega}}| \gg \tau_\eta^{-1}$, allowing us to use approximate solution \cite{PhysRevB.79.045302}
\begin{align}
    S_z(t) &\approx e^{-(\bm{r}-{\rm \textbf{v}} t)^2/2w^2}e^{-t/\tau_s}\cos(\omega t),
    \\
    \dfrac{1}{\tau_s} &= \dfrac{1}{2}
    \left(
    \dfrac{1}{\tau_z}
    +
    \dfrac{\sin^2\theta}{\tau_x}
    +
    \dfrac{\cos^2\theta}{\tau_y}
    \right)
\end{align}
with $\omega = g\mu_B B_{\rm ext}/\hbar$, and $\theta$ defined by the direction of the total field $\bm{B}_{\rm tot} = \bm{B}_{SO} + \bm{B}_{\rm ext}$. Since $B_{SO} \propto {\rm v}$, one could expect $\theta$ to change as ${\rm v}$ increases, leading to the anisotropy in $\tau_s$. However, since $B_{SO} \ll B_{\rm ext}$ as shown above, the angle is either $\theta \approx 0$ or $\pi/2$ for $\bm{B}_{\rm ext} \parallel \hat{y}$ or $\hat{x}$, respectively. Therefore one should not expect an anisotropy induced by ${\rm v}$. Nevertheless, we can estimate the maximum anistropy that could be reached with this model using the SOC extracted from the experimental data, which yield $\Delta \tau_s \sim 10$\%, much smaller than the observed anisotropy shown in Fig.~\ref{fig5}.

The drift corrections to $\tau_\eta
^{-1}$ are also negligible, since they scale as $({\rm v}/v_F)^2 \ll 1$ \cite{PhysRevB.95.125119}, where the Fermi velocity $v_F \sim 250$~km/s is large due to the large Fermi energy $\varepsilon_F/k_B \sim 150$~K (in temperature units). The drift velocity could lead to Joule heating, which would affect the thermal averages and enhance $\beta_3 \propto \mean{k^2}$ \cite{PhysRevB.86.174301,Kunihashi2017,PhysRevB.99.125404,aipadvances2020}, which in turn would affect the DP relaxation rates $\tau_\eta^{-1}$. However, while we estimate a heating from 10 K to $\sim$60 K at the largest drift velocities, for large $\varepsilon_F \gg k_BT$ the heating effects on $\beta_3$ are negligible \cite{PhysRevB.86.174301,Kunihashi2017,PhysRevB.99.125404,aipadvances2020}.

The overall large spin lifetime $\tau_S$, and particularly the peak seen in Fig. \ref{fig3}(h') resembles the reported PSH formation when the Dresselhaus and Rashba SOCs are matched by electrically tuning $\alpha$ \cite{Kohda2012GatePSH, Yoshizumi2016GatePSHInv, Kunihashi2016, Dettwiler2017}. However, due to our large pump spot size ($\sim 23$ $\mu$m) one would expect that spatial correlations vanish and the uniform packet ($k=0$) dynamics should follow \cite{Salis2014PSHLocalExc}, which does not excite the finite $k$ PSH modes. Moreover, due to the lack of a complete model for such PSH formation in coupled two-subband systems, we cannot confirm the origin of the lifetime enhancement.

These estimates shown above do not match the long spin lifetime, strong anisotropy and drift response seen in the experimental data. We believe that the source of the discrepancy is that current two-subband models \cite{FuPRL, PhysRevB.95.125119} neglect intersubband correlations, which might be relevant in nearly-degenerate cases, as in our experimental data. For instance, in Ref.~\cite{PRB2016} the extremely small subband splitting (0.14 meV) allowed us to rotate the subbands from symmetric (S) and anti-symmetric (AS) configurations into left (L) and right (R) wells eingestates. This transformation interchanges the intra- and inter-subband Rashba coupling. While on the S/AS basis the Rashba coupling is zero, on the L/R basis it is large. Here the 2 meV subband splitting is not small enough to allow for this rotation, and the intersubband SOC is expected to be small perturbative corrections \cite{FuPRL}. Nonetheless, if intersubband correlations play a significant role, one might need a more complete spin diffusion model that accounts for intersubband SOC \cite{Bernardes, Calsaverini, fu2015} beyond perturbative corrections to the subband dispersions.

\section{Conclusions}
Long coherence time is vital for quantum processing technologies. In a spin transistor, it will limit the capability to deliver information at the drain location. We investigated a device containing a 2DEG hosted in a wide QW with two-subbands occupied. The effect of in-plane and gate voltages on the spin relaxation time was studied using a pump-probe method. The data showed very peculiar features depending on the transport direction. For drift velocities along $\left[110\right]$, the lifetime was strongly reduced and remain constant independently of the applied gate voltage. On the contrary, for transport along $\left[1\bar{1}0\right]$, the spin lifetime was longer than in the $\left[110\right]$ and gate-dependent regardless of the in-plane voltage. Such strong anisotropy and drift velocity dependence of $\tau_s$ and $|\Omega|$ cannot be explained by the available two-subband theoretical models. We have considered these models, and possible extensions for finite temperature, but it shows weak anisotropy and nearly velocity independent $\tau_s$. Therefore a more detailed model for the two-subband problem is still needed, and may require the inclusion of intersubband SOC \cite{Bernardes,Calsaverini,fu2015} and intersubband correlations. The exceptionally large values of $\tau_s \sim 7$~ns and the strong anisotropy are excellent motivations for these theoretical extensions and for more experimental studies in multisubband systems.

\begin{acknowledgments}
F.G.G.H acknowledges financial support from the São Paulo Research Foundation (FAPESP) Grants No. 2015/16191-5, No. 2016/50018-1, and No. 2018/06142-5, and Grant No. 301258/2017-1 of the National Council for Scientific and Technological Development (CNPq). G.J.F. acknowledges the financial support from CNPq and FAPEMIG. V.S. was supported by the U.S. Department of Energy, Office of Basic Energy Sciences, Division of Materials Sciences and Engineering, Award DE-SC0016206.
\end{acknowledgments}

\bibliography{PRBlifetime}
\end{document}